\documentclass[a4paper,11pt]{scrartcl}

\usepackage[utf8x]{inputenc}	
\usepackage{fullpage}
\usepackage{fourier} 
\usepackage[english]{babel}		

\usepackage{FiraSans}
\usepackage{soul}

\usepackage{amsmath,amstext,amssymb,amsfonts,amsthm,mathrsfs,chemarrow,stackrel,dsfont}

\usepackage{xcolor}
\usepackage{graphicx} 					
\usepackage{subcaption}					

\usepackage[sort&compress]{natbib} 

\usepackage{microtype}
\usepackage{array}
\usepackage{multirow}
\usepackage{colortbl}
\usepackage{booktabs}

\usepackage[pdftex, breaklinks, colorlinks, citecolor=black, urlcolor=black, linkcolor=black]{hyperref} 

\usepackage[toc,page,header]{appendix}
\usepackage{minitoc}
\usepackage{chapterbib}

\usepackage{titlesec}
\titleformat{\section}[block]{\Large\bfseries}{\thesection}{1em}{}
\titleformat{\subsection}[block]{\large\bfseries}{\thesubsection}{1em}{}
\titleformat{\subsubsection}[block]{\large\itshape}{\thesubsubsection}{1em}{}
\titleformat{\paragraph}[runin]{\itshape}{\theparagraph}{1em}{}[. ]

\allowdisplaybreaks[2]

\newcolumntype{P}[1]{>{\centering\arraybackslash}m{#1}}

\definecolor{darkgreen}{rgb}{0.0, 0.42, 0.24}
\definecolor{darkblue}{rgb}{0.0, 0.0, 0.55}
\definecolor{Gray}{gray}{0.75}
\definecolor{change}{rgb}{0.5,0.,0.6}

\title{The stochastic dynamics of early epidemics: probability of establishment, initial growth rate, and infection cluster size at first detection}
\author{Peter Czuppon$^{1,2,3}$, Emmanuel Schertzer$^{4}$, Fran\c{c}ois Blanquart$^{2,5,\ast}$, \\ Florence D\'{e}barre$^{1,\ast}$}
\date{}

\begin{document}

\doparttoc 
\faketableofcontents 

\renewcommand \thepart{}
\renewcommand \partname{}

\maketitle

$^1$ Institute of Ecology and Environmental Sciences of Paris (iEES-Paris, UMR 7618), Sorbonne Universit\'e, CNRS, UPEC, IRD, INRAE, 75252 Paris, France \vspace{3pt}

$^2$ Center for Interdisciplinary Research in Biology, CNRS, Coll\`ege de France, PSL Research University, 75005 Paris, France \vspace{3pt}

$^3$ Institute for Evolution and Biodiversity, University of M\"{u}nster, 48149 M\"{u}nster, Germany \vspace{3pt}

$^4$ Faculty of Mathematics, University of Vienna, 1090 Wien, Austria
\vspace{3pt}

$^5$ Universit\'{e} de Paris, INSERM, IAME, 75018 Paris, France \vspace{3pt}

$^\ast$ equal contributions \vspace{7pt}

\clearpage
%

\clearpage


\section{Introduction}

The emergence and spread of infectious diseases pose an increasing threat in an ever more interconnected world.
%
A quantitative understanding of epidemic dynamics is necessary to improve control measures. 
Deterministic models are a suitable tool to describe the epidemiological dynamics once a large number of individuals has been infected.
During the early phase of an epidemic or a local infection cluster however, stochastic effects cannot be neglected. These stochastic effects are due to the initially low number of infected individuals, and to the inherent stochasticity of the transmission process. Understanding and quantifying these stochastic effects will help, for example, assess the risk of new infection clusters emerging or estimate the size of a cluster associated with a new variant when such a variant is detected.

The infectiousness of an individual may vary over the course of their infection because of within-host viral dynamics if the transmission rate is correlated with the viral load.
We consider a generic stochastic model in which infectiousness is an arbitrary function of time since infection.
This stochastic model is called a Crump-Mode-Jagers process \citep{crump_68,crump_69,jagers_69}. When the number of infected individuals gets large, this stochastic model can be approximated by a deterministic partial differential equation describing the distribution of the time since infection of the host population. This equation is known as the McKendrick-von Foerster partial differential equation \citep{mckendrick_1925, diekmann_book, foutel-rodier_2020}.
 
Transmission timings are particularly influential during the early stages of the growth of an infection cluster, which is the focus of our work. It is therefore important to use biologically realistic distributions of transmission times~\citep{linton_2020}, rather than assuming mathematically convenient but biologically unrealistic exponential distributions. A constant infectiousness over the duration of an individual's infection leads to the predominantly used framework of ordinary differential equations, while non-constant infectiousness can be captured by a partial differential equation. 
In addition to the added biological realism, a time-varying infectiousness of infected individuals can also properly capture the dynamical consequences of abrupt changes in transmission rate \citep{foutel-rodier_2020,forien_2021}. This is not possible with an ordinary differential equation framework~\citep{gatto_2020}.

Here, we provide key results about the epidemic dynamics as described by the McKendrick-von Foerster equation.  
Stochasticity in transmission does not merely add noise to the dynamics, but also causes a systematic deviation from the deterministic description, which underestimates the initial growth of an establishing epidemic \citep{mercer2011, rebuli2018}. This is in contradiction to a common misconception that stochasticity generally slows down the initial epidemic growth rate. We quantify the deviation between the deterministic and observed stochastic growth rate by conditioning the individual-based process on survival. After initial stochastic effects, the process converges to exponential growth with an asymptotic growth rate, denoted $r$, derived from the reproduction number $R$ and the transmission rate. The distribution of time since infection in the stationary regime is exponential with parameter $r$, the asymptotic growth rate.

The reviewed and newly derived results can provide answers to public health related questions: How many importations will eventually result in a local infection cluster? How large is a local cluster once a first case is detected? When did a new variant -- like Alpha, first detected in the United Kingdom (UK) -- arise? How large is the detection rate of infectious individuals by a single mass testing effort? How many daily tests need to be conducted to detect local clusters before they exceed a certain size? We show how our theoretical results provide quantitative answers to these questions.


\section{Expected epidemic size}
%
We study the epidemic size of a cluster initiated by a single  infected individual. We refer to a 'cluster' as the entire tree of infections that was initiated by a single infected individual. In particular, we do not spatially restrict a cluster, nor do we constrain the time period in which transmissions need to occur.

Because some of our developments will also need them, we first recall results on \textit{deterministic} epidemiological dynamics, then develop new analytical results on the expected early growth and the expected number of infected individuals once a stationary regime has been reached. We illustrate with simulations the variability across stochastic trajectories (Fig.~\ref{fig:init_growth}). As observed before \citep{mercer2011, rebuli2018}, the expected growth rate during the early phase of cluster growth is greater than the long-term deterministic expectation, because clusters that do not die out are typically those that initially grow faster. We show how to account for this phenomenon in the mathematical description of the early phase and of the stationary regime.

In our stochastic simulations, we assume that the epidemic starts with a single infected individual at time $t=0$. Each infected individual $i$ is assigned a time since infection $a_i$.  
The time since infection determines the infectiousness of an individual through time. The term `time since infection' is also referred to as `age of infection' in the mathematical literature. We decouple the transmission rate $\tau(a)$ into a mean number of secondary infections $R$ and a transmission probability density over time $\mu(a)$. We then have
\begin{equation}
    \tau(a) = R \times \mu(a)\, .
\end{equation}
This equation holds because $\int_0^\infty \mu(a)da=1$, so that indeed the average number of secondary infections is given by $R$.
This decoupling allows us, in a relatively simple way, to study different offspring distributions for $R$, while leaving the transmission density $\mu(a)$ unchanged.

For illustration, we assume that the distribution of transmission times follows a gamma distribution, but any distribution would be possible. In particular, a constant transmission rate (uniform distribution) would result in an exponential distribution of the transmission times (i.e., the memory-less distribution), which would reduce this general model into an ordinary differential equation (ODE).

\subsection{Previous results on deterministic dynamics: renewal equation, growth rate and time-since-infection distribution}
Throughout our analysis, we assume that the fraction of susceptible individuals is sufficiently large compared to the number of individuals infected in the early epidemic that it remains approximately constant.
The overall rate at which new infections occur at time $t$, denoted by $i(t)$, in the deterministic regime is described by the following renewal equation \citep{wallinga2006}:
\begin{equation}\label{eq:renewal_incidence}
    i(t) \ = \ \tau(t) \ + \  \int_0^t \tau(a) i(t-a) da\, ,
\end{equation}
where $\tau(a)$ is the transmission rate of an individual with time since infection $a$. The first term $\tau(t)$ reflects the new infections by the first infected individual at time $t$. The integral in Eq.~\eqref{eq:renewal_incidence} is the continuous version of the sum over the number of new infections caused by individuals with time since infection $a$ (term $i(t-a)da$), which happens at rate $\tau(a)$. Intuitively, one can think about $i(t)dt$ being the incidence at time $t$, i.e., the number of newly infected individuals in the small time interval $[t,t+dt]$.

The cumulative number of infected individuals, i.e., the total epidemic size, which we denote by $I(t)$, is then given by 
\begin{equation}\label{eq:renewal}
    I(t) = 1 + \int_0^t i(s) ds\, = 1 + \int_0^t I(t-a) \tau(a) da\, ,
\end{equation}
with $I(0)=1$ (mathematical details are given in the Supplementary Information (SI), Section~S4). 
For simplicity, we do not consider recovery of infected individuals. However, individuals will of course stop transmitting when the time since infection is such that the transmission rate $\tau(a)$ becomes very small. 

The epidemic size $I(t)$ will, for large times $t$, grow exponentially if $R>1$. Formally, the asymptotic exponential growth rate $r$ is obtained by solving the classical Euler-Lotka equation \citep{wallinga2006,britton2019}:
\begin{equation}\label{eq:lotka-euler}
    1 = \int_0^\infty e^{-rt} \tau(t)dt \quad \Leftrightarrow \quad \frac{1}{R} = \int_0^\infty e^{-rt}\mu(t)dt,
\end{equation}
where $r$ is also called the Malthusian parameter of the supercritical branching process \citep{haccou_book}. In the case where $\mu(t)$ is given as the density of a gamma distribution with shape parameter $\alpha$ and scale parameter $\beta$, the exponential growth rate $r$ is
\begin{equation}\label{eq:lotka-euler_gamma}
    r = \frac{R^{1/\alpha}-1}{\beta}\, .    
\end{equation}

Convergence speed from the initial condition towards the asymptotic growth rate $r$ is determined by the average number of secondary infections $R$ and the transmission probability density $\mu$. Intuitively, the faster a large number of infected individuals is reached (high $R$ and/or small average transmission time), the faster is convergence towards the stationary growth regime. 

Furthermore, it is possible to derive an explicit expression for the number of infected individuals over time, once asymptotic growth is reached. It follows from results of supercritical general branching processes and renewal theory \citep{haccou_book}, that the expected cumulative epidemic size is, for asymptotically large times $t$, given by 
\begin{equation}\label{eq:det_dynamics}
    I(t) = I(0) \frac{e^{r t}}{r R \int_0^\infty e^{-r s} s \mu(s) ds}\ .
\end{equation}
The integral in the denominator is the mean generation time of the Malthusian process \citep{svensson2007,britton2019}. This is the time between the infection of the infecting individual and the time of infection of a randomly chosen secondary infection event. If the transmission density $\mu(s)$ were constant, the integral would be $1/(rR)$ and the epidemic size would be the solution of a constant infection process without depletion of susceptibles: $I(t) = I(0) e^{r t}$. 

For an uncontrolled COVID-19-epidemic (we set $R=2.9$, estimated for the French epidemic in Spring 2020~\citep{salje_2020}), we obtain $r\approx 0.18$ per day, which corresponds to a doubling time of about $4$~days. 
When interventions are in place (e.g., $R=1.3$), then the Malthusian parameter is $r\approx 0.048$ per day, which corresponds to a doubling time of $14$~days.

Under exponential growth, the distribution of the ages of infection in the population is given by an exponential distribution with parameter $r$, the exponential growth rate \citep{haccou_book}. Intuitively, in an exponentially growing population, the number of individuals who were infected $a$ days ago is $e^{r}$ times greater than the number of individuals who were infected $a+1$ days ago. 
The exponential distribution also implies that for a large growth rate $r$, a large proportion of the cumulative number of infections will be very recent. For example, with $R=2.9$, $30\%$ among the total cumulative number of infections occurred within the last two days. 

We now turn to the stochastic simulations and show how systematic deviations from the deterministic regime can be understood and mathematically described.
We first give a stochastic correction for the asymptotic growth rate and then apply a similar idea to the general epidemic size process over time.

\subsection{Asymptotic growth rate and epidemic size in the stochastic epidemic model}
For large enough times after the initially infected individual started the local cluster, the epidemic grows exponentially at the rate predicted by the Euler-Lotka equation (Eq.~\eqref{eq:lotka-euler}). 
However, the expected cumulative epidemic size derived for the deterministic case (Eq.~\eqref{eq:det_dynamics}) includes epidemics that eventually die out. Since we are only interested in epidemic clusters that eventually result in a large epidemic outbreak, we rescale the initial epidemic size by dividing by the survival probability~$p_{\text{surv}}$:

\begin{equation}
    \label{eq:rescaled_det_dynamics}
    I_{\text{surv}}(t) = \frac{I(t)}{p_{\text{surv}}} =  \frac{I(0)}{p_{\text{surv}}}\ \frac{e^{r t}}{r R \int_0^\infty e^{-r s} s \mu(s) ds}\ .
\end{equation}
This rescaling reflects conditioning of the epidemic process on survival (Fig.~\ref{fig:init_growth}). The survival probability is~$p_{\text{surv}} = 1-p_{\text{ext}}$, where the probability of extinction $p_\text{ext}$ is numerically computed as the fixed point of the probability generating function of the distribution of secondary infections. In words, the probability of extinction is equal to the probability that the initial infected individual does not produce any secondary infection, plus the probability that it produces one secondary infection which goes extinct ($p_\text{ext}$), plus the probability that it produces two secondary infections which both go extinct ($p_\text{ext}^2$), and so on; this intuition is outlined in SI, Section~S1.
Formally, the correction of the asymptotic limit in Eq.~\eqref{eq:rescaled_det_dynamics} is derived from a convergence result of a general branching process (SI, Section~S3). 

\subsection{Initial stochastic growth of an epidemic}
\begin{figure}[t]
    \centering
  	\includegraphics[width=0.6\textwidth]{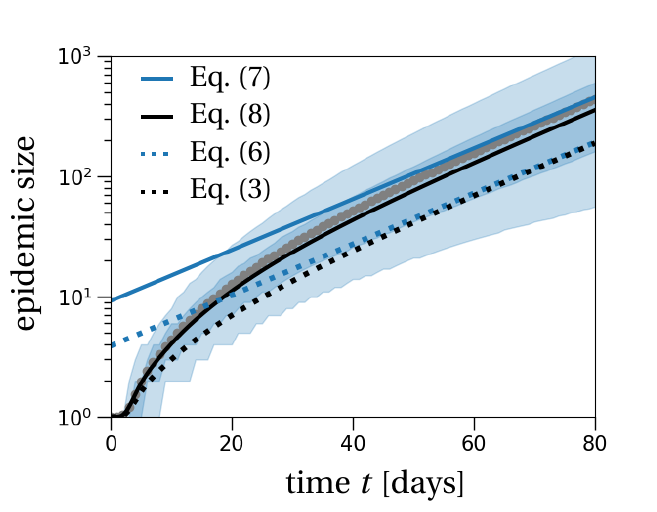}
  	\caption{\textbf{The cumulative number of infected individuals over time.} The light and dark shaded regions show the 90\% and 50\% inter-quantile ranges obtained from 10,000 stochastic simulations that resulted in cluster establishment. Gray dots show the average of these simulations over time. The theoretical prediction (black solid line) is calculated from Eq.~\eqref{eq:epi_size_adj} with the adjusted transmission rate as computed in SI, Section~S5. The black dotted line shows the prediction obtained from Eq.~\eqref{eq:renewal} without the conditioning for the epidemic to establish. The solid blue line is the epidemic size predicted by the asymptotic growth rate as stated in Eq.~\eqref{eq:rescaled_det_dynamics}. The blue dotted line is the corresponding quantity without the stochastic adjustment (Eq.~\eqref{eq:det_dynamics}). The effective reproduction number is set to $R=1.3$, the number of secondary transmission events is Poisson-distributed, and the transmission density $\mu(t)$ is a Gamma distribution with the parameters given in Table~\ref{tab:variant_params}.}
    \label{fig:init_growth}
\end{figure}

The initial growth rate of an epidemic that does not become extinct is initially steeper than its final asymptotic growth rate \citep{mercer2011, rebuli2018} (compare the initial slope of the mean of stochastic simulations with the asymptotic growth for large times; gray dots vs. blue solid line in Fig.~\ref{fig:init_growth}).
This is due to the inherent stochasticity of the transmission process, which strongly affects the dynamics when there are only a small number of infected individuals. Clusters that escape extinctions are typically those that by chance benefited from a larger initial growth than the long-term expectation.
This also means that deterministic models tend to underestimate epidemic sizes early on, or, if parameters are inferred from data, overestimate epidemic parameters such as the true basic reproduction number $R_0$, as for example observed in \citep{kochaczyk2020}.

To account for this initial stochastic phase, one can alter the individual-based dynamics by conditioning the stochastic process on the survival of the epidemic. A similar procedure has been employed in \citep{rebuli2018}. This conditioning results in an adjustment of the transmission rate $\tau$, which we denote by $\widetilde{\tau}$. Formally, this adjustment is only justified for the stochastic process by Doob's h-transform \citep{doob_1957} (details in SI, Section~S5). In the large population size limit, we then approximate the adjusted transmission rate by the continuous analog of the adjusted transmission rate of the stochastic process. This approximation, while mathematically not fully justified, is a natural analogy of the conditioning of the asymptotic epidemic size in Eq.~\eqref{eq:rescaled_det_dynamics}. The mean epidemic size of the adjusted process is then computed by 
\begin{equation}\label{eq:epi_size_adj}
    \widetilde{I}(t) = 1 + \int_0^t \widetilde{i}(s) ds\, ,
\end{equation}
where $\widetilde{i}(s)ds$ is the incidence in the time interval $[s,s+ds)$ under the adjusted process. 
The rate of new infections $\widetilde{i}(t)$ in the conditioned process now depends non-linearly on the history of the epidemic and therefore does not satisfy a renewal equation as in Eq.~\eqref{eq:renewal_incidence}, but a delay differential equation:
\begin{equation}\label{eq:ren_adj_incidence}
    \widetilde{i}(t) = F(\widetilde{i}(s); s\in [0,t])\, .
\end{equation}
The function $F$ is explicitly computed in SI, Section~S5 (Eq.~(S37)). In short, the conditioning on survival of the epidemic results in an adjustment of the transmission rate $\tau$ by a factor that varies over time. This adjustment factor reflects the survival probability of the epidemic at a certain time and depends on the size and the age structure of the epidemic over time. The adjustment factor is largest at time $t=0$, where it equals $(1+p_{\text{ext}})$. Over time, the adjustment factor decreases and asymptotically approaches $1$ for a large epidemic size, where the probability of extinction becomes negligible, i.e., for large times $\widetilde{\tau}=\tau$. 

In Fig.~\ref{fig:init_growth}, we plot both the adjusted and non-adjusted versions of the mean epidemic size (Eqs.~\eqref{eq:renewal} and~\eqref{eq:epi_size_adj}). As mentioned above, the non-adjusted formula (black dotted line) underestimates the mean epidemic sizes as obtained from 10,000 stochastic simulations (gray dots). In contrast, conditioning the transmission density on survival (black solid line) predicts the mean epidemic size over time reasonably well, and also equilibrates approximately at the correct level. Overall, there is large variation in the epidemic sizes between different trajectories, as shown by the broad light shaded region corresponding to the 90\% inter-quantile range of the simulated trajectories. To model the number of secondary infections, we have used the Poisson distribution in the figure because the adjustment of the transmission rate does not result in explicit expressions for the negative binomial case. Cumulative epidemic sizes in case the number of secondary infections is distributed according to a negative binomial or geometric distribution show more variation due to the larger variance in the number of secondary infections (Fig.~S2 in SI, Section~S6). 

\section{Applications}
We now apply the theoretical results obtained above. First, we use the approximation of the epidemic size (Eq.~\eqref{eq:epi_size_adj}) to estimate the probability distribution of the emergence time of the Alpha variant, first detected in the UK in September 2020. The distribution of the emergence time also provides insight into the probability distribution of the size of the cluster when the variant was first sampled. As a second application, we estimate the minimal testing frequency necessary to detect new emerging clusters before they exceed a certain size (on average). This prediction is especially relevant when the number of infected individuals is rare.

\subsection{Distribution of the first detection time and cluster size at detection, and application to the origin of the Alpha variant}
The Alpha variant initially consisted only of the B.1.1.7 lineage. This lineage was first detected in the UK from a sample that was collected on September 20$^\text{th}$ 2020~\citep{rambaut_blog} and has rapidly become a major variant of concern due to its increased transmissibility~\citep{volz_2021} and pathogenicity~\citep{davies2021}. Here, we develop a method to estimate the first infection of an individual with the Alpha variant and the distribution of the size of the Alpha-cluster on the day when the sample was taken in September, based on the dynamics of the epidemic size of a local cluster. 

Our analysis requires the effective reproduction number, estimated to be $R=1.5$ for the Alpha variant in November 2020 in the UK~\citep{volz_2021}, and the probability for a sample taken in the UK to be sequenced, which was around $4.2\%$ in October 2020~\citep{sampling}. We will use this value in our analysis, keeping in mind that this might be an underestimate because the number of cases has been lower in September so the percentage of samples that could have been sequenced is potentially higher. Since only reported cases can be sampled, we additionally account for underreporting of cases. We assume that around 25\% of all infections are detected~\citep{colman2021}. Lastly, we need to define a distribution for the time that passes between infection and sampling of an infectious individual. We assume that the time from infection to sampling is a gamma distributed random variable (but any distribution would work) with a mean of seven days and a standard deviation of two days. The parameter values (Table~\ref{tab:variant_params}) are chosen such that they give a probability of sampling an infected individual up until 3 days of their infection that is less than 1\%, and a probability of sampling an infected individual after ten days of their infection that is less than 10\%.
All parameters are summarised in Table~\ref{tab:variant_params}.

\begin{table}[tb]
    \centering
    \begin{tabular}{P{3.5cm}|P{3cm}|P{3cm}|P{3cm}}
        \textbf{Interpretation} & \textbf{Distribution} & \textbf{Parameters} & \textbf{Reference} \\
        \hline \hline
            \rowcolor{Gray} mean number of secondary infections & Poisson & $R = 1.5$ & \citep{volz_2021} \\
            
            time of secondary infection & Gamma \qquad (density: $\mu(t)$) & shape: 6.6, \qquad scale: 0.833 \qquad (mean: 5.5 days) &      \citep{hinch2021}\\
            
            \rowcolor{Gray} time from infection to sampling & Gamma \qquad (density: $f_{\text{sampling}}(t)$) & shape: 12,\qquad scale: 7/12 \qquad (mean: 7 days) & -- \\
            
            sequencing probability & Bernoulli & $p_{\text{sequencing}} = 0.042$ & \citep{sampling} \\
            \rowcolor{Gray} sampling probability & Bernoulli & $p_{\text{sampling}} = 0.25 \times p_{\text{sequencing}}$  & \citep{colman2021} \\ 
        \bottomrule
    \end{tabular}
    \caption{\textbf{Probability distributions and parameter values used in the case study of the Alpha variant.}}
    \label{tab:variant_params}
\end{table}

\subsubsection*{Distribution of the first detection time}
To estimate the time of the first detection of an individual infected by the Alpha variant, we combine the sampling probability distribution $f_{\text{sampling}}$ with the expected epidemic size at time $t$, given by the adjusted version of the epidemic size in Eq.~\eqref{eq:epi_size_adj}, and the number of infections until the first infected in the cluster is sampled and sequenced, which happens with probability $p_{\text{sampling}}$ per infected individual. For readability, we refer to this first infected individual that is sampled and sequenced by case X and only write sampling when in fact we mean sampling and sequencing.
The number of infection events till case X is infected, including case X, is denoted $N_{\text{inf}}$. It is a geometrically distributed number with probability $p_{\text{sampling}}$. 
Note that if we were interested in the $j^{\text{th}}$ sampling event, the number of infected individuals until the $j^{\text{th}}$ sampling event would be distributed according to a negative binomial distribution with `success' probability $p_{\text{sampling}}$ and dispersion $\kappa = j$. 

We combine the distribution of $N_{\text{inf}}$ with the \textit{deterministic} time needed for the infected population to reach $N_{\text{inf}}$ individuals (conditioned on non-extinction of this epidemic cluster as computed in Eq.~\eqref{eq:epi_size_adj}). We also refer to this time as \textit{hitting time} and denote it by $t_{N_{\text{inf}}}^{\text{det}}$. To this, we add the time from infection of case X to their sampling.
Denoting by $T_{\text{sampling}}$ the random variable corresponding to the time of first detection and sampling, its probability density is given by:
\begin{equation}\label{eq:samp_time_dist}
\begin{aligned}
    h_{\text{sampling}}(t) &:= \lim_{\text{dt}\to 0}\mathbf{P}\left(T_{\text{sampling}} \in (t-\text{dt},t+\text{dt})\right) \\
    &\approx \sum_{i=1}^{\infty} \mathbf{P}\left(N_{\text{inf}}=i\right)\  f_{\text{sampling}}\left(t-t_i^{\text{det}}\right) \\
    &= \sum_{i=1}^\infty p_{\text{sampling}} \left(1-p_{\text{sampling}}\right)^{i-1} f_{\text{sampling}}\left(t-t_i^{\text{det}}\right) \, ,
\end{aligned}
\end{equation}
where $f_{\text{sampling}}(s)$ denotes the probability density of the time from infection to sampling evaluated at time $s$ (Table~\ref{tab:variant_params}). We emphasize that the density of the first sampling time $h_{\text{sampling}}(t)$ is an approximation, because it is based on the mean epidemic size and not the whole distribution of the epidemic size. The mean epidemic size directly provides the deterministic hitting time $t_i^{\text{det}}$, neglecting the whole distribution of the epidemic size. 

\begin{figure}[t]
    \centering
	\includegraphics[width=.7\linewidth]{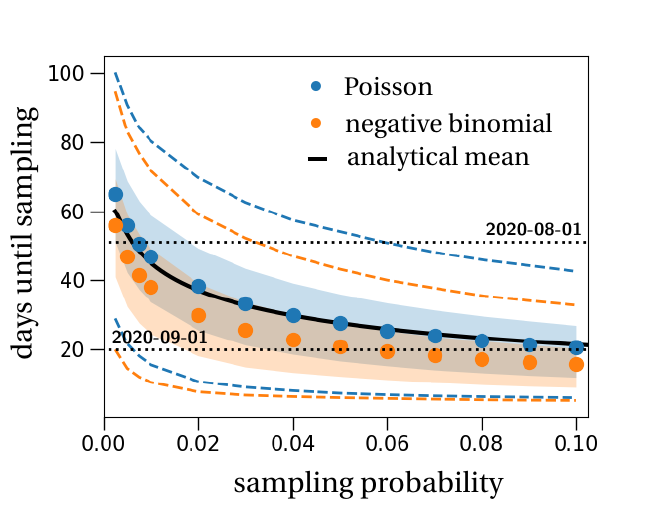}
  	\caption{\textbf{The date of emergence of the first infected with the Alpha variant in the UK when varying the sampling probability.} The shaded regions and dashed lines show the 50\% and 90\% inter-quantile ranges obtained from 10,000 stochastic simulations that resulted in cluster establishment; blue for the secondary infections being Poisson distributed, orange for a negative binomial distribution. Dots represent the means of these simulations when varying the sampling probability. The effective reproduction number is set to $R=1.5$, the dispersion parameter is $\kappa=0.57$~\citep{salje_2020}, and the transmission density $\mu(t)$ and the waiting time between infection and sampling ($f_{\text{sampling}}$) are Gamma distributions with parameters as stated in Table~\ref{tab:variant_params}. The theoretical mean (black solid line) of the first sampling time is calculated from Eq.~\eqref{eq:samp_time_dist}, which only applies to the Poisson case.} 
    \label{fig:date}
\end{figure}

With our COVID-19-specific parameter set given in Table~\ref{tab:variant_params}, we find that the mean time between the first infection of an individual with the Alpha variant and sampling of case X is around 46~days, indicating that the strain was present in the UK the 4$^{\text{th}}$ of August 2020 -- yet, the variance is quite large for this distribution: the standard deviation is 19.5~days. The emergence date of the Alpha variant strongly depends on the sampling probability: smaller sampling probabilities result in earlier possible emergence dates than larger probabilities (Fig.~\ref{fig:date}).
The distribution of secondary cases also impacts the timing: if the number of secondary infections is distributed as a negative binomial distribution, the date of emergence shifts closer to the date of sampling of case X. This effect is secondary though, compared to the impact of the sampling probability (Fig.~\ref{fig:date}). 

\begin{figure}[t]
    \begin{subfigure}{.5\textwidth}
  		\centering
  		\includegraphics[width=\linewidth]{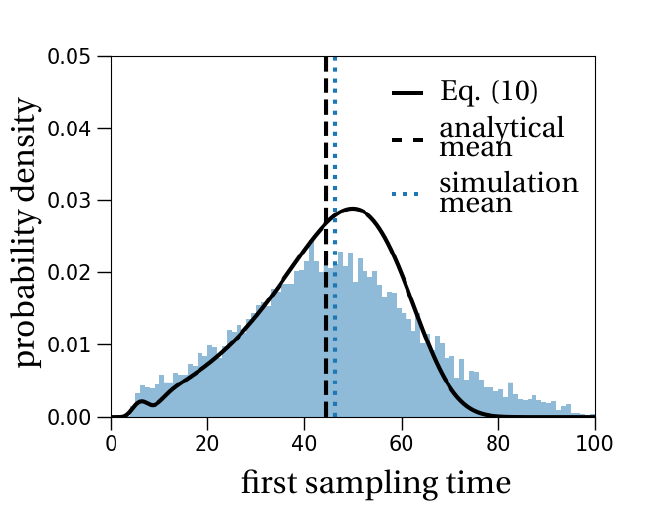}
  		\caption{first sampling time}
	\end{subfigure}%
	\begin{subfigure}{.5\textwidth}
 		 \centering
 		 \includegraphics[width=\linewidth]{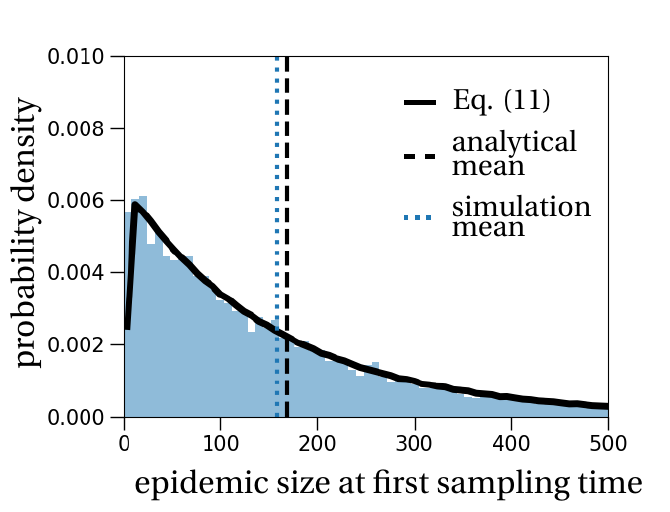}
  		\caption{cluster size at first sampling}
	\end{subfigure}
	\caption{\textbf{Distribution of the first sampling time and the cluster size at that time, parameterized to the case of the Alpha variant.} The histograms are obtained from 10,000 stochastic simulations and represent (a) the first sampling time of an infected individual with the Alpha variant, measured since the first infection of an individual with the Alpha variant (in days), and (b) the cluster size at this first sampling time. The theoretical predictions (black solid lines) are computed by Eqs.~\eqref{eq:samp_time_dist} and~\eqref{eq:size_dist}. The parameters and distributions used in the stochastic simulations are given in Table~\ref{tab:variant_params}.}
    \label{fig:variant}
\end{figure}

In general, we find that the theoretical prediction of the probability distribution of the first sampling time captures the shape of the empirical distribution from the stochastic simulation results (Fig.~\ref{fig:variant}a). Note that this implies that most of the variability in time does not come from stochasticity in epidemic size, but from the variability emerging from the random sampling of infected individuals ($p_{\text{sampling}}$) and the variability in the time from infection to sampling of infected individuals ($f_{\text{sampling}}$). Biologically, the variability in the time from infection to sampling arises from inter-individual variability in viral dynamics, symptom development, test seeking behaviour, etc. 
We find the largest discrepancy between theory and simulations at large first sampling times, i.e., we underestimate the right tail of the first sampling time distribution. This difference arises because our theoretical approximation does not take into account variability in the epidemic size process. Fig.~\ref{fig:init_growth} shows a large variation in the number of infected individuals over time between different stochastic trajectories. Most notably, there are several trajectories that remain at low cumulative epidemic sizes for a relatively long time. These trajectories are responsible for the long right tail of the sampling time distribution in Fig.~\ref{fig:variant}a.

\subsubsection*{Cluster size at the first detection time}
Next, we use this distribution of the first sampling time to infer the size of the epidemic cluster at that time. Therefore, we combine the adjusted epidemic size in Eq.~\eqref{eq:epi_size_adj} with Eq.~\eqref{eq:samp_time_dist} and obtain the following probability mass function for the size of the cluster at the sampling time of case X:
\begin{equation}\label{eq:size_dist}
    \mathbf{P}\left(I(T_{\text{sampling}}) = k\right) = \int_0^\infty h_{\text{sampling}}(t) \mathds{1}_{\{ \widetilde{I}(t)\in [k-1/2,k+1/2)\}} dt\, ,
\end{equation}
where $h_{\text{sampling}}$ is the probability that the first sampling time lies in the interval $[t,t+dt)$, given in Eq.~\eqref{eq:samp_time_dist}.

This estimate of the epidemic size distribution approximates the simulated data reasonably well (Fig.~\ref{fig:variant}b). The only notable difference occurs for very low epidemic sizes, where the epidemic size at the first sampling time ranges from 0-8 (bin size is set to 8 -- the smallest bin size that produces a continuous theoretical prediction), as can be seen in the histogram in Fig.~\ref{fig:variant}b. The mean size of the cluster with the Alpha variant at the first sampling time (obtained from stochastic simulations) consists of 159~individuals, yet again with a large standard deviation of 158~individuals. For example, the 95-percentile of the simulations predicts a cluster size of 476~infected individuals with the Alpha variant by the time of the first sampling of the variant.

\subsection{Minimal testing frequency to detect clusters of a given size}
A single mass testing effort only results in a detection rate of between 25-48\% of potentially infectious individuals, depending on the utilized test (rapid test or polymerase chain reaction) and the exponential growth rate $r$ corresponding to reproduction numbers $R$ between 1.3 and 3 (details in SI, Section~S7). Therefore, we now ask whether repeated random testing in the population is a more feasible strategy to contain an infection cluster. Specifically, how often should we randomly test the population to detect a cluster before it exceeds a certain size? As a numerical example we will use a threshold cluster size of 30 infected individuals.  
We assume that testing is applied population-wide at random, independently of the infection state of an individual. 
The probability to test positive depends on the time since infection of an individual~\citep{borremans2020,kucirka2020,Hellewell2021}. 
We denote the probability to test positive by a rapid test if the infected individual has been infected $a$ days ago by $Q(a)$ (Fig.~S3 in SI, Section~S7). 

If a fraction $f$ of the population is tested every day, the detection probability of an infected individual is approximately given by 
\begin{equation}\label{eq:det_prob_testing}
    p_{\text{detect}} = 1 - \prod_{a=1}^\infty (1 - f Q(a)) \approx f \sum_{a=1}^\infty Q(a)\, .
\end{equation}
The term $(1 - f Q(a))$ is the probability that an infected individual is not detected at their time since infection $a$. Hence, the product is the probability that an individual is never detected over the course of the individual's infection. The probability of detection is one minus this product. The approximation is valid when it is very unlikely that the same individual is tested more than once during the period when there is a high chance to detect their infection.

To determine the testing frequency above which the expected cluster size is smaller than 30 infected individuals, we repeat the steps from the previous sections: first, we determine the first detection time and then translate this result to the average cluster size at detection. Since our analytical result tends to overestimate the cluster size at detection (Fig.~S4 in SI, Section S8), this analytical procedure will provide an upper bound for the true testing frequency required to detect clusters of a certain size. In our numerical example with $R=1.1$, this procedure results in a testing frequency of 0.013 for a threshold cluster size of 30 infected individuals. 

Importantly, increasing the testing frequency when it is still low offers large benefits in terms of cluster size at detection because the epidemic size at detection reflects the exponential growth of the epidemic: it decreases exponentially with increasing testing frequency (Fig.~S4 in SI, Section S8).

\section{Discussion}
We have collected key equations and derived novel results to account for stochasticity during the early phase of epidemic trajectories.
Explicitly taking into account stochastic effects during the early phase of an epidemic allowed us to compute a good description of the mean epidemic size for all times (Eq.~\eqref{eq:epi_size_adj}). Importantly, our result captures the increased initial growth rate of surviving epidemics when compared to the asymptotic growth rate (Fig.~\ref{fig:init_growth}). This is a known effect \citep{mercer2011,rebuli2018}, yet cannot be captured by deterministic epidemiological models. 
One important consequence of this theoretical underestimation of classically used models is that parameter inference during the early phase of an epidemic of, for example, the basic reproduction number, will result in an overestimation of the true value~\citep{kochaczyk2020}. We provide a new mathematical description of the expected epidemic size over time that could be used in statistical inference during the early phase of emerging epidemics.

As a first application, we analytically derived the probability distribution of the first detection time of an epidemic cluster. While in principle applicable to any type of detection event, as for instance the first death or the first hospitalization event, we have focused on dating the emergence of the Alpha variant that was first sampled in the UK the 20$^{\text{th}}$ of September 2020. 
Our analysis is appropriate for clusters that descend from a single infected individual, and as long as population immunity is low enough for the supply of susceptible individuals to be unlimited. The Alpha variant was first detected in England in September 2020 and likely emerged there once, so our analysis can be applied to it. It would not apply, for example, to the Delta variant in the UK, unless the cluster linked to the first importation of the variant could be identified -- and so the date of importation could be estimated. 
On average, we find that the Alpha cluster was started 46 days before its detection, which means that the variant was likely present in the UK on the 4$^{\text{th}}$ of August 2020. Usually, phylogenetic methods are used to date the evolutionary history of mutations \citep{hadfield2018}. 
In this particular case, a phylogenetic approach is difficult because of the large divergence between Alpha and non-Alpha variants sampled at a similar time~\citep{rambaut_blog}. Indeed, we did not find a published estimate of the date of emergence of the Alpha variant based on a phylogenetic analysis. 
In an attempt to date the origin of SARS-CoV-2, a combination of phylogenetic and epidemiological methods has been used to obtain a more complete picture of the very early dynamics of the COVID-19 epidemic~\citep{pekar2021}. Our new description for early epidemic growth provides a formal non-spatial description of the individual-based simulations that were used in \citep{pekar2021} to date the very first COVID-19 case.

We additionally derived an analytical approximation for the probability distribution of the epidemic size at the first detection event. In contrast to a previous numerical estimate of the cluster size at the first disease-caused death that relies solely on the waiting time distribution until detection, e.g. the distribution from infection to death~\citep{jombart_2020}, we consider the whole epidemic trajectory of the cluster, i.e., from the first infected individual to the day of detection. The previously proposed method inevitably results in an overestimate of the actual epidemic size. Previous research has also shown that if the probability of detection since infection were constant over time, which is not the case in our setting, the cluster size at detection would be geometrically distributed~\citep{trapman2009,lambert2013}. Whether the distribution of cluster sizes at detection is a geometric distribution if the detection process is not constant in time, is an open question. In our specific data set, this seems to be the case (Fig.~\ref{fig:variant}b). 

We also applied our results to the evaluation of testing strategies.
Currently (May 2021), aside from vaccination campaigns, frequent testing is seen as a possible solution to relax COVID-19-related restrictions in the short term. Our modelling approach gives an estimation of the minimal testing frequency per day to detect epidemic clusters of a certain size, for example small enough for manual contact tracing to be feasible. The minimal testing frequency depends on the test that is employed. In our numerical example, we have used the detection probability estimated for rapid tests, which were collected during the early phase of the epidemic in the UK in 2020 \citep{Hellewell2021}. Since then, tests have improved so that our estimation of the minimal testing frequency is very likely an overestimate. We find that for a cluster size to be below 30 infected individuals (on average), each day around 0.13\% of a total population would need to be randomly selected for testing, i.e., independently of the individual's infection status. 
Pooled sample testing strategies could be a solution to reduce the number of testing kits needed, and is a particularly reasonable option when the prevalence of infected individuals in a community is close to zero~\citep{brault2021}.

Additionally, we estimated the fraction of cases that can be detected during a single mass testing effort, as has been for example conducted in Slovakia in fall 2020 \citep{pavelka2021}.
We find that with either a rapid test or a polymerase chain reaction test and with a reproduction number between $R=1.3$ and $R=2.9$, the detection rate of infectious individuals is between 26-48\% (SI, Section~S7). 
During the mass testing effort, a certain fraction of undetected individuals is still in the latent phase (0-3 days post-infection) and will become infectious after the mass testing event. 
Similar observations have also been made by employing a deterministic SEIR-model \citep{bosetti2021}. 
This indicates that only isolating positively detected individuals would be insufficient to contain the epidemic and that mass testing would need to be repeated to efficiently control the epidemic. \\

In conclusion, we have summarised existing theoretical results describing the early, stochastic dynamics of an epidemic, and developed new results on the mean epidemic size trajectory. We combined the establishment probability with the deterministic McKendrick-von Foerster equation to obtain a precise description of the expected epidemic size of an establishing epidemic over time. As an application, we approximated the probability distribution for the timing of a first infected individual in an epidemic cluster. This distribution can be used to estimate, for example, the emergence of new variants of a pathogen, like the Alpha variant. In addition, we derived the minimal testing frequency to detect clusters below a certain size. These applications are relevant from a public health perspective and could be used to guide the policy to contain and fight any infectious disease.

\subsection*{Data availability}
The C++ codes, data files and Python scripts used to generate the figures are available at \url{https://gitlab.com/pczuppon/early-epidemic-dynamics}. 

\subsection*{Funding}
PC has received funding from the European Union's Horizon 2020 research and innovation program under the Marie Sk{\l}odowska-Curie grant agreement PolyPath 844369. FD is funded by an Agence Nationale de la Recherche JCJC grant TheoGeneDrive ANR-19-CE45-0009-01. FB is funded by a Momentum grant from the CNRS.




\bibliographystyle{mystyle}
\bibliography{covid.bib}

\renewcommand{\theequation}{S\arabic{equation}}
\renewcommand{\thetable}{S\arabic{table}}
\renewcommand{\thesection}{S\arabic{section}}
\renewcommand{\thefigure}{S\arabic{figure}}
\setcounter{equation}{0}  
\setcounter{figure}{0}
\setcounter{table}{0}


\clearpage
\begin{appendix}
\section{Probability of establishment}\label{sec:estab}

Introductions of infected individuals into a susceptible population do not always result in a local infection cluster because of random extinction events. Here, we quantify the probability that the introduction of a single infected individual results in the establishment of a local cluster.
The establishment probability of a local cluster depends on the probability distribution of the number of secondary infections, which also determines the chance of superspreading events. 
Similarly, the initial spread of new variants can be modeled just as emerging clusters. 
In the following, we briefly outline how to compute the probability of establishment and study how it is affected by different types of transmission distributions.

In our epidemiological model, we assume that the number of susceptible individuals is not limiting the spread of the disease, i.e., that the fraction of susceptible individuals in the population remains close to one. 
Every infected individual transmits the disease to $R$ other individuals on average, where $R$ is the effective reproduction number. The actual number of secondary infections can vary strongly between infected individuals. For example, estimates for COVID-19 indicate that about 20\% of infected individuals are responsible for about 80\% of secondary infections~\citep{endo_2020}. These superspreaders (or superspreading events) cannot be captured by a Poisson-distributed number of secondary infections~\citep{lloyd-smith2005}. A more dispersed distribution, i.e. with a larger variance, is the negative binomial distribution, where most of infected individuals do not transmit the disease at all. Its variance is typically quantified by the dispersion parameter $\kappa > 0$. The smaller the value of $\kappa$, the more variance has the negative binomial distribution. 

The establishment probability of a local cluster can be computed by following the transmission chain from one generation of infected individuals to the next. Specifically, if $Y$ is the random number of secondary infections due to one infected individual, the probability of extinction, i.e., 1 minus the probability of survival, is computed by summing over the possible numbers of secondary infections times the probability that all corresponding chains of transmission do not survive:
\begin{equation}\label{eq:pgf}
    p_{\text{ext}} = \underbrace{\mathbf{P}(Y=0)}_{\substack{\text{no secondary}\\ \text{infection}}} + \underbrace{p_{\text{ext}}\mathbf{P}(Y=1)}_{\substack{\text{one secondary}\\ \text{infection}}} + \underbrace{p_{\text{ext}}^2 \mathbf{P}(Y=2)}_{\substack{\text{two secondary}\\ \text{infections}}} + p_{\text{ext}}^3 \mathbf{P}(Y=3) + \dots = \mathbf{E}[p_{\text{ext}}^Y]\, .
\end{equation}
The expectation on the right, $\mathbf{E}[z^Y]$, is called the probability generating function and can be computed for several distributions explicitly. Eq.~\eqref{eq:pgf} then shows that the probability of extinction $p_{\text{ext}}$ of a cluster started with a single infected individual, is given by the smallest positive fixed point of the probability generating function, i.e., $\mathbf{E}[z^Y] = z$ \citep{haccou_book}.
If the epidemic is started with $k$ infected individuals, the extinction probability is simply given by $p_{\text{ext}}^k$, which yields a survival probability of $p_{\text{surv}}=1-p_{\text{ext}}^k$. Here, and also in the first equality of Eq.~\eqref{eq:pgf}, we have used the assumption that the transmission chains of the $k$ initially infected individuals are independent of each other.  

Plugging in different distributions for the number of secondary infections, we can numerically compute the probability of extinction. In the particular case of a geometric distribution, which is a negative binomial distribution with dispersion parameter $\kappa=1$ (compare to Eq.~\eqref{eq:distributionsNBP} below), the fixed point equation can be solved analytically by
\begin{equation}\label{eq:geom_ext}
    \mathbf{E}[z^Y] = \frac{p}{1-(1-p)z} \quad \Rightarrow \quad p_{\text{ext}}=\frac{p}{1-p} = \frac{1}{R}, 
\end{equation}
where $p$ is the success probability of the geometric distribution and $R$ is the average number of secondary infections (or effective reproduction number). The last equality in Eq.~\eqref{eq:geom_ext} is obtained by noting that $R$ is equal to the mean of the distribution, and that the mean of a geometric distribution is $(1-p)/p$.  

The probability generating functions for the negative binomial and the Poisson distributions are given by
\begin{equation}\label{eq:distributionsNBP}
    \mathbf{E}[z^Y] = \left\{ 
    \begin{array}{ll}
        \left(\frac{p}{1-(1-p)z}\right)^\kappa, & \begin{array}{l} \hspace{-5pt}\text{neg. binomial dist. with mean $R$ and dispersion parameter $\kappa$} \\ 
        \hspace{-5pt}\text{(success probability } p = \frac{\kappa}{\kappa + R})\, ,\end{array} \\
        e^{R(z-1)}, & \text{Poisson distribution with mean } R\, .
    \end{array}
    \right.
\end{equation}
Note that the Poisson distribution is obtained from the negative binomial distribution in the limit $\kappa\to\infty$. The fixed point of the probability generating functions of a negative binomial or a Poisson distribution cannot be computed analytically. 

\begin{figure}[t]
    \centering
  	 \begin{subfigure}{.5\textwidth}
  		\centering
  		\includegraphics[width=\linewidth]{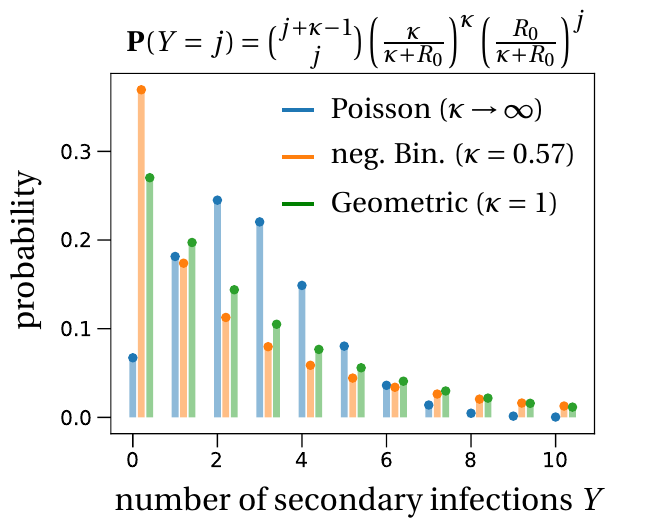}
  		\caption{Distribution of secondary infections}
  		\label{fig:offspring}
	\end{subfigure}%
	\begin{subfigure}{.4775\textwidth}
 		 \centering
 		 \includegraphics[width=\linewidth]{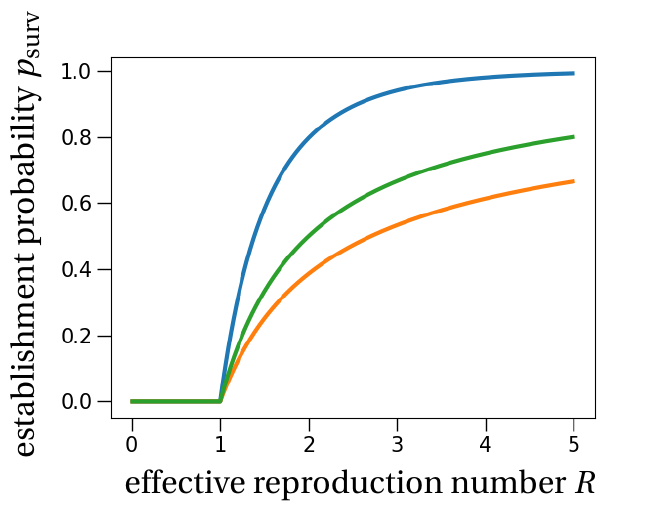}
  		 \caption{Establishment probability}
  		 \label{fig:establishment}
	\end{subfigure}
  	\caption{\textbf{Probability distribution and establishment probability of an epidemic for different distributions of the number of secondary infections.} (a) The probability for an infected individual to transmit the disease to $Y$ susceptible individuals depends on the distribution of the number of secondary infections. The probability to not transmit the disease ($Y=0$) is highest for the negative binomial distribution with $\kappa=0.57$ (orange bars; value estimated for the French COVID-19 data pre-lockdown 2020; \citet{salje_2020}), i.e., the distribution with the highest variance. The probability of establishment is smallest for the offspring distribution with the largest variance, as shown in (b). For $\kappa=1$, the geometric distribution (green curve), the establishment probability is explicit: $p_{\text{surv}}=1-1/R$. The largest establishment probability is found for a Poisson-distributed number of secondary transmission events (blue curve).
  	}
    \label{fig:estab}
\end{figure}

In Fig.~\ref{fig:estab}, we plot the distributions and establishment probabilities as a function of $R$ for the three different distributions of number of secondary infections. For $R$ smaller than 1, the epidemic will not establish, so that the establishment probability is 0. For values of the effective reproduction number R greater than 1, the establishment probability becomes positive and the epidemic has a chance to establish. In general, the smaller the offspring variance, i.e., the larger the dispersion parameter $\kappa$ of the negative binomial distribution, the larger the probability of establishment. In applications, the analytically exact result $1/R$ (Eq.\eqref{eq:geom_ext}) is often used to approximate the extinction probability of an epidemic. Fig.~\ref{fig:estab} shows that this is indeed a reasonably good approximation for overdispersed offspring distributions such as the negative binomial distribution with a parameter $\kappa<1$, at least as long as $R$ does not become too large.

\clearpage
\section{Time-since-infection dependent probability of extinction}
\label{app:ext_prob}
In this section, we compute the extinction probability of the epidemic when started with a single infected individual with time since infection $a$ (also called 'age of infection'). In the previous section, we have only considered the special case $a=0$. 

We consider a branching process approximation of the epidemic. We assume that each individual is characterized by its time since infection and we set $\tau(a)$ as the mean infectiousness at age $a$, i.e.,
\begin{equation}
    \tau(a) da \ = \ \mathbf{P}( \mbox{individual with time since infection in $[a, a + da]$ infects a new individual} ).
\end{equation}

For $a>0$, let $p_{\text{ext}}(a)$ be the probability that the epidemic starting with a single individual with time since infection $a$ does not establish (in the previous section we have computed $p_{\text{ext}}(0)$). More formally, if $Z_t$ denotes the number of individuals infected at time $t$, then 
\begin{equation} \label{eq:ext_prob}
    p_{\text{ext}}(a) \ = \ \mathbf{P}_a\bigg( \limsup Z_t <\infty \bigg) = \sum_{k=0}^\infty p_{\text{ext}}(0)^k \mathbf{P}(Y_a = k), 
\end{equation} 
where $\mathbf{P}_a$ is the probability distribution of the branching process starting from a single individual with age $a$ and $Y_a$ denotes the number of future secondary infections of an individual with age $a$.
From this, in the case of a Poisson-distributed number of secondary infections, we find
\begin{equation}\label{eq:ext_prob_deriv}
    \frac{d}{da} p_{\text{ext}}(a) \ = \ \tau(a) \left(1-p_{\text{ext}}(0)\right) p_{\text{ext}}(a).
\end{equation}
As a consequence
\begin{equation}\label{eq:ext_prob_age}
     p_{\text{ext}}(a) \ = \  p_{\text{ext}}(0) \exp\bigg( (1-p_{\text{ext}}(0)) \int_0^a \tau(s) ds \bigg)\, ,
\end{equation} 
where $p_{\text{ext}}(0)$ can be determined by the boundary condition
\begin{equation} 
    \lim_{a\to\infty} p_{\text{ext}}(a) = 1\, , 
\end{equation}
which is ensured by the fact that $R<\infty$. (Intuitively, an individual infected a long time ago is not infectious anymore.) We then find $p_{\text{ext}}(0)$ as the unique solution of 
\begin{equation}\label{eq:fixed} 
    p_{\text{ext}}(0) \exp\bigg( (1-p_{\text{ext}}(0)) R \bigg) = 1, \ \text{where} \ p_{\text{ext}} (0)<1.  
\end{equation}
Note that the expression of $p_{\text{ext}}(0)$ is the same as the one stated the previous section in the case of a Poisson-distributed number of secondary infections.

\bigskip

\noindent

\clearpage
\section{Derivation of the asymptotic limit}
\label{app:convergence}
We consider a Crump-Mode-Jagers process, also referred to as a general branching process, with an unspecified distribution of secondary infections to model the number of infected individuals over time. The random variable for the number of secondary infections is denoted by $Y$. 
We define $\tau$ as the intensity measure of the process, i.e., such that for every test function $f$
\begin{equation}
    \mathbf{E}\left[\sum_{i=1}^Y f(t_i)\right] \ = \ \int_0^\infty f(u)\tau(u) du  
\end{equation}
and the correlation intensity measure $c$ such that
\begin{equation}
    \mathbf{E}\left[ \mathds{1}_{Y\geq 2} \sum_{i\neq j}  f(t_i,t_j) \right] \ = \ \int_0^\infty f(u,v)c(u,v) du dv \, .  
\end{equation}
We assume the Kesten-Stigum condition 
\begin{equation}
    \int_0^\infty u \ln(u) \tau(u) du <\infty \, .
\end{equation} 
Under this assumption, there exists a random variably $W_\infty$ such that
\begin{equation}\label{eq:w_infty}
    \exp(-r t) Z_t \ \to \ W_\infty  \ \ \mbox{almost surely,}
\end{equation}
where $r$ is the Malthusian parameter as defined in Eq.~(4) in the main text, and $Z_t$ is the number of infected individuals at time $t$.

Let $A$ be the non-extinction event. We know that $\mathbf{E}[W_\infty]=1$ and that  almost surely $A = \{W_\infty >0\}$. 
It follows that
\begin{equation}\label{eq:w_infty_con}
    \mathbf{E}[W_\infty | A ] \ = \ \frac{1}{1-p_{\text{ext}}}\, ,
\end{equation}
where $p_{\text{ext}}$ is the extinction probability, which is the smallest root of the probability generating function as defined in Eq.~\eqref{eq:pgf} in the main text:
\begin{equation}
    p_{\text{ext}} = \mathbf{E}[ p_{\text{ext}}^{Y} ]. 
\end{equation}
Combining Eqs.~\eqref{eq:w_infty} and~\eqref{eq:w_infty_con} gives the formal justification for the correction of the initial epidemic size by $1/(1-p_{\text{ext}})$ in Eq.~(7) in the main text.

Further, if $\psi(t)=\mathbf{E}[e^{tY}]$ is the Laplace transform (or moment-generating function) of $W_\infty$, then $\psi$ satisfies the fixed point problem \citep[][Theorem 4.1]{Cohn1985}
\begin{equation}
    \mbox{for all } u\geq 0,\ \  \psi(u) \ = \ \mathbf{E}\bigg[ \prod_{i=1}^{Y} \psi(u e^{-r t_i})  \bigg].
\end{equation}

Let us now evaluate the variance of $W_\infty$ conditional on non-extinction.
Differentiating the Laplace transform $\psi$ twice with respect to $u$, we get 
\begin{eqnarray*}
    \psi''(u) & = & \mathbf{E}\bigg[\mathds{1}_{Y\geq 1} \  \sum_{i}   e^{-2r t_i}  \psi''(u e^{-r t_i}) \prod_{j\neq i}   \psi(u e^{-r t_j}) \bigg]  \ + \ \\
    & & \mathbf{E}\bigg[ \mathds{1}_{Y\geq 2} \  \sum_{i\neq j}   e^{-r t_i} e^{-r t_j}   \psi'(u e^{-r t_i}) \psi'(u e^{-r t_j}) \prod_{k\neq i,j}   \psi(u e^{-r t_k}) \bigg]\, .
\end{eqnarray*}
Evaluating at $u=0$, we find
\begin{eqnarray*}
    \mathbf{E}[W_\infty^2] &  = &  \mathbf{E}[W_\infty^2] \mathbf{E}\bigg[ \mathds{1}_{Y\geq 1} \  \sum_{i}   e^{-2r t_i}   \bigg]  \ + \   \psi'(0)^2 \mathbf{E}\bigg[ \mathds{1}_{Y\geq 2} \  \sum_{i\neq j}   e^{-r t_i} e^{-r t_j}  \bigg]\, .
\end{eqnarray*}
Since $\psi'(0) \ = \ 1/(1-p_{\text{ext}})$, this yields
\begin{eqnarray*}
    \mathbf{E}[W_\infty^2]  \ =  \  \frac{1}{(1-p_{\text{ext}})^2}\frac{ \mathbf{E}\bigg[ \mathds{1}_{Y\geq 2} \  \sum_{i\neq j}   e^{-r t_i} e^{-r t_j}  \bigg]}{1-  \mathbf{E}\bigg[ \mathds{1}_{Y\geq 1} \  \sum_{i}   e^{-2r t_i}   \bigg] }\, .
\end{eqnarray*}
With $A = \{W_\infty >0\}$ as above, which coincides with the event of non-extinction, we find 
\begin{eqnarray*}
    \mathbf{E}[W_\infty^2 | A]  & =  &  \left(\frac{1}{1-p_{\text{ext}}}\right)\frac{ \mathbf{E}\bigg[ \mathds{1}_{Y\geq 2} \  \sum_{i\neq j}   e^{-r t_i} e^{-r t_j}  \bigg]}{1-  \mathbf{E}\bigg[ \mathds{1}_{Y\geq 1} \  \sum_{i}   e^{-2r t_i}   \bigg] }  \\
    & = & \left(\frac{1}{1-p_{\text{ext}}}\right) \  \frac{ \int_0^\infty \exp(-r (t+u)) c(t,u) dt du }{1-\int_0^\infty  \exp(-2r t) \tau(t) dt }.
\end{eqnarray*}
Then
\begin{eqnarray*}
    \mbox{\textbf{Var}}[W_\infty^2 | A]  & = & \frac{1}{1-p_{\text{ext}}} \bigg( \frac{ \int_0^\infty \exp(-r (t+u)) c(t,u) dt du }{1-\int_0^\infty \exp(-2r t) \tau(t) dt } - \frac{1}{1-p_{\text{ext}}}\bigg).
\end{eqnarray*}
 \clearpage

\section{Renewal equation in the absence of conditioning}
\label{app:renewal}
We informally derive the renewal equation of the incidence and the epidemic size as given in Eqs.~(2) and~(3) in the main text.

In the following, $\sigma_x$ denotes the time of infection of individual $x$ and
${\cal F}_t$ denotes the history of the infection process up to time $t$. (In the probabilistic jargon, $({\cal F}_t)_{t\geq 0}$ is the natural filtration associated with the infection process).
Let us define the random empirical measure, which counts the number of infections in a small time interval $[t,t+dt]$
\begin{equation}
    d b(t) \ = \ \sum_{x: \sigma_x <\infty} \delta_{\sigma_x}(dt),
\end{equation}
where $\delta_{\sigma_x}$ is the Dirac measure. 
The random empirical measure $db$ records the infection times along the course of the epidemic. Each atom of the random measure $db$ corresponds to a time of infection.
Note that since the epidemic is triggered by a single individual with time since infection (or 'age') $0$ at time $0$, 
$db$ has a Dirac mass at $0$. For $t>0$, we can assume without loss of generality that $\mathbf{E}[d b]$
has no atom and we write 
\begin{equation}\label{eq:i_def}
    \text{for all } t>0, \ \ i(t) := \frac{\mathbf{E}[db(t)]}{dt}\, ,
\end{equation}
where $i(t)$ is the overall rate at which new individuals are infected in a small time interval $[t,t+dt]$. In other words, $i(t)dt$ models the incidence in the time interval $[t,t+dt]$.

Our goal is to show that $i$ satisfies the renewal equation
\begin{equation}\label{eq:i-renewal}
i(t) \ = \ \tau(t) \ + \  \int_0^t \tau(t-u) i(u) du\, . 
\end{equation}
Informally, we have
\begin{equation}
\mathbf{E}\bigg[ d b(t) \ \vert \ {\cal F}_t \bigg] \ =  \sum_{ \sigma_x <t }   dt \times \tau(t-\sigma_x)\, .
\end{equation}
In words, if we condition on the history of the process up to time $t$, the expected number of new infections in the time interval $[t,t+dt]$ can be computed by adding up the expected contribution of every infected individual before time $t$. For individual $x$, this expected contribution is $dt \times \tau(t-\sigma_x)$ and the previous equation follows by adding up the contribution of every previously infected individual.

Next, the sum on the right hand side of the latter equation can be rewritten in terms of an integral with respect to the infection measure $d b$ in the following way
\begin{equation}
\sum_{ \sigma_x <t }   dt \times \tau(t-\sigma_x) \ = \ dt \int_{[0,t)} \tau(t-u) d b(u)\, ,
\end{equation}
so that 
\begin{equation}
\mathbf{E}\bigg[ d b(t) \ \vert \ {\cal F}_t \bigg] \ =  \ dt \int_{[0,t)} \tau(t-u) d b(u)\, .
\end{equation}
Averaging over every realization of the process up to time $t$,  we obtain
\begin{equation}\label{eq:lin}
\begin{aligned}
\mathbf{E}\bigg[ d b(t) \bigg] & = dt\ \mathbf{E}\bigg[\int_{[0,t)} \tau(t-u) d b(u) \bigg] \\
& =\ dt \int_{[0,t)} \tau(t-u) \mathbf{E}[d b(u)]\, .
\end{aligned}
\end{equation}
Recalling our definition of $i(t)$ in Eq.~\eqref{eq:i_def}, this yields
\begin{equation}
i(t) \ = \ \tau(t) \ + \  \int_0^t \tau(t-u) i(u) du\, ,
\end{equation}
where the term $\tau(t)$ on the right hand side follows from the fact that $d b$ has a Dirac mass at $0$. This is Eq.~(2) in the main text.

The cumulative epidemic size $I(t):=1+\int_0^t i(s)ds$ then satisfies the renewal equation
\begin{equation}
    \begin{aligned}
        I(t)&=1 + \int_0^t i(s)ds \\
        & = 1 + \int_0^t \big(\tau(s)+ \int _0^s i(s-u)\tau(u)du\big) ds \\
        & = 1+ \int_0^t \tau(s) ds  + \int_0^t \tau(u) \int_u^t i(s-u) ds\ du \\ 
        & = 1 + \int_0^t \tau(s) ds  + \int_0^t \tau(u)  \underbrace{\int_0^{t-u} i(s) ds}_{=I(t-u)-1} du \\
        & = 1 + \int_0^t \tau(a) I(t-u) du\, , 
    \end{aligned}
\end{equation} 
where we have used Fubini's theorem to change the order of integration. This explains Eq.~(3) in the main text.

\clearpage
\section{Delay differential equation for the process conditioned on explosion}
\label{app:delay}

In this section, we derive a delay differential equation for the rate of new infections over time when the process is conditioned on non-extinction. This delay differential equation explains Eq.~(9) in the main text.

\bigskip
\noindent{\textbf{Markov model.}}  For simplicity, we assume that the infection process is Markovian in its age structure, where by age structure we refer to the time since infection composition in the infected population. More precisely, this means that the times since infection for a given individual are given by a Poisson point process with intensity $\tau$. At time $t>0$, we encode the population by a random vector ${\mathbf a} = (a_1, a_2,\cdots, a_n)$
where $a_1>\cdots>a_n$ are the times since infection listed in decreasing order (individual $1$ is the first infected individual etc.), and $n$ is the number of infected individuals at time $t$.
We denote by $|{\mathbf a}|$ the number of elements in ${\mathbf a}$.
The infinitesimal generator of the infection process is then given by
\begin{equation} 
Lf ({\mathbf a}) \ = \ \sum_{i=1}^{|{\mathbf a}|}  \left[\tau(a_i) \bigg( f(({\mathbf a},0)) - f ({\mathbf a}) \bigg) \ + \ \partial_{i} f(\mathbf a) \right]\, ,
\end{equation}
where $({\mathbf a},0):= (a_1,\cdots, a_{|{\mathbf a}|},0)$ and $f$ is a continuous and bounded test function \citep[e.g][]{ethier_book}. Intuitively, one can think of the infinitesimal generator to be the expected rate of change of a Markov process, which generalizes the concept of the transition rate matrix in the case of a discrete state continuous time Markov chain.
It encodes the infinitesimal dynamics of the branching process in the following way: on a time interval $dt$, 
\begin{itemize}
    \item an individual
with time since infection $a_i$ produces a new infection with a probability $\tau(a_i) dt$. Conditional on this event, the infection age (time since infection) structure of the population makes a transition from ${\mathbf a}$ to $({\mathbf a},0)$, i.e., an individual with time since infection $0$ is added to the population.
\item the time since infection of every individual increases by $dt$ (this corresponds the partial derivative of $f$ in the generator). 
\end{itemize}    

\bigskip 
\noindent{\textbf{The Doob $h$-transform.}} Let us now consider the probability of survival, which is equivalent to explosion ($Z_t\to\infty$ for $t\to\infty$),
\begin{equation} h({\mathbf a})  \ := \ 1- \prod _{i=1}^{|{\mathbf a}|} p_{\text{ext}}(a_i).\end{equation}
for a population with initial time-since-infection structure ${\mathbf a}$. 
Applying Eq.~\eqref{eq:ext_prob_deriv}, we note that $h$ is a harmonic function for the generator $L$, i.e., $L h=0$.
Let us consider the Markov process with infinitesimal generator  
\begin{equation} \widetilde L f = \frac{1}{h} L (f h)\, .  \end{equation}
This is the Doob $h$-transform of the original process.

\bigskip

\noindent{\textbf{Branching process conditioned on explosion.}} The Markov process with generator $\widetilde L$ is known to be identical in law with the original process conditioned on explosion~\citep{doob_1957,chetrite_2014}.
Using $Lh=0$, the transformed infinitesimal generator computes to
\begin{equation} 
\begin{aligned}
 \widetilde L f(\mathbf a)  & =   \sum_{i=1}^{|{\mathbf a}|}  \left[\tau(a_i) \frac{h(({\mathbf a},0))}{h({\mathbf a})} \bigg( f(({\mathbf a},0)) - f ({\mathbf a}) \bigg) \ + \ \partial_{i} f(\mathbf a) \right]   \\
 & = \sum_{i=1}^{|{\mathbf a}|}  \left[\tau(a_i) \frac{ 1- p_{\text{ext}}(0)\prod _{i=1}^{|{\mathbf a}|} p_{\text{ext}}(a_i)  }{ 1- \prod _{i=1}^{|{\mathbf a}|} p_{\text{ext}}(a_i)  } \bigg( f(({\mathbf a},0)) - f ({\mathbf a}) \bigg) \ + \ \partial_{i} f(\mathbf a) \right]\, . 
\end{aligned}
\end{equation} 
The formula can now be interpreted as follows. For a given time-since-infection structure ${\mathbf a}$, conditioning on explosion induces an adjustment of the time since infection of the $i^{\text{th}}$ individual as follows
\begin{equation}\label{eq:trafo}
\widetilde \tau^{{\mathbf a}}(a_i) \ = \  \tau(a_i)  \ \cdot \ \underbrace{\frac{ 1- p_{\text{ext}}(0)\prod _{j=1}^{|{\mathbf a}|} p_{\text{ext}}(a_j)  }{ 1- \prod _{j=1}^{|{\mathbf a}|} p_{\text{ext}}(a_j)  } }_{\mbox{adjustment}}. 
\end{equation}
We note that the infection rate is increased for every individual, independent of their time since infection, by the same amount. The increase, however, depends on the time-since-infection structure of the entire infected population. 

\bigskip
\noindent\textbf{{The delay equation.}} We keep the same notation as in Section~\ref{app:renewal}. By the same reasoning as in Section~\ref{app:renewal}, we have
\begin{equation}\label{tau}
\mathbf{E}[ d b(t)  | {\cal F}_t ] \ =  \sum_{ \sigma_x <t }  \bigg( \frac{1-p_{\text{ext}}(0) \prod_{y:\sigma_y<t} p_{\text{ext}}(t-\sigma_y) }{ 1- \prod_{y:\sigma_y<t} p_{\text{ext}}(t-\sigma_y) }  \bigg) \ dt \times \tau(t-\sigma_x)\, . 
\end{equation}
In words, Eq.~\eqref{tau} means that if we condition on the history of the process up to time $t$, the expected number of new infections in the time interval $[t,t+dt]$ can be computed by adding up the expected contribution of every infected individual before time $t$.
According to Doob's $h$-transform formula,
for an individual infected at time $\sigma_x$, the expected contribution of this individual is given by $ dt \times \tau(t-\sigma_x)$ multiplied by the Doob's term 
\begin{equation} 
\bigg( \frac{1-p_{\text{ext}}(0) \prod_{y:\sigma_y<t} p_{\text{ext}}(t-\sigma_y) }{ 1- \prod_{y:\sigma_y<t} p_{\text{ext}}(t-\sigma_y) }  \bigg)\, ,
\end{equation}
which depends on the whole time-since-infection structure of the population up to time $t$. 

We now rewrite the adjustment factor in the following way
\begin{equation}
    \begin{aligned}
        \frac{1-p_{\text{ext}}(0) \prod_{y:\sigma_y<t} p_{\text{ext}}(t-\sigma_y) }{ 1- \prod_{y:\sigma_y<t} p_{\text{ext}}(t-\sigma_y) }  & = \ 
        \frac{1-p_{\text{ext}}(0)\exp\bigg( \sum_{y:\sigma_y<t}\ln(  p_{\text{ext}}(t-\sigma_y)) \bigg) }{ \exp\bigg(\sum_{y:\sigma_y<t} \ln( p_{\text{ext}}(t-\sigma_y))\bigg) } \\
        & = \  \frac{1-p_{\text{ext}}(0)\exp\bigg(  \int_{[0,t)} \ln( p_{\text{ext}}(t-u) ) d b(u)   \bigg) }{ 1- \exp\bigg(\ \int_{[0,t)} \ln( p_{\text{ext}}(t-u) ) d b(u)\bigg) }\, .
\end{aligned}
\end{equation} 
As a consequence, Eq.~\eqref{tau} can be rewritten as 
\begin{equation}
\begin{aligned}
\mathbf{E}[ d b(t)  | {\cal F}_t ] 
& = \ dt\  \bigg(\sum_{x: \sigma_x <t}  \tau(t-\sigma_x) \bigg)  \ \times 
\frac{1-p_{\text{ext}}(0)\exp\bigg(  \int_{[0,t)} \ln( p_{\text{ext}}(t-u) ) db(u)   \bigg) }{ 1- \exp\bigg( \int_{[0,t)} \ln( p_{\text{ext}}(t-u) ) db(u) \bigg) }  \\
& = \ dt \int_{[0,t)}  \tau(t-u) d b(u)  \ \times \ 
\frac{1-p_{\text{ext}}(0)\exp\bigg(  \int_{[0,t)} \ln( p_{\text{ext}}(t-u) ) d b(u)   \bigg) }{ 1- \exp\bigg( \int_{[0,t)} \ln( p_{\text{ext}}(t-u) ) db(u) \bigg)  } \, .
\end{aligned}
\end{equation}
Averaging over the past of the infection process ${\cal F}_t$, we find that
\begin{equation}
\begin{aligned}
\mathbf{E}[ d b(t) ] 
& = & dt\ \mathbf{E}\bigg[\int_{[0,t)}  \tau(t-a) d b(a)  \ \times \ 
\frac{1-p_{\text{ext}}(0)\exp\bigg(  \int_{[0,t)} \ln( p_{\text{ext}}(t-a) ) d b(a)   \bigg) }{ 1- \exp\bigg( \int_{[0,t)} \ln( p_{\text{ext}}(t-a) ) db(a) \bigg)  } \bigg].  \label{eq:i-cond}
\end{aligned}
\end{equation}
Since $db$ has a Dirac at $0$ we have
\begin{equation}\begin{aligned}
\mathbf{E}[ d b(t) ] 
& = & dt \mathbf{E}\bigg[\ \bigg(\tau(t)+\int_0^t  \tau(t-u) d b(u) \bigg)   \ \times \ 
\frac{1-p_{\text{ext}}(0) p_{\text{ext}}(t)\exp\bigg(  \int_0^t \ln( p_{\text{ext}}(t-u) ) d b(u)   \bigg) }{ 1- p_{\text{ext}}(t)\exp\bigg( \int_0^t \ln( p_{\text{ext}}(t-u) ) db(u) \bigg)  } \bigg].  \label{eq:ii-cond}
\end{aligned}
\end{equation}

Let us now compare this formula with its analog in Eq.~\eqref{eq:lin} in the absence of conditioning the process on survival. 
In Eq. \eqref{eq:lin}, we can use Fubini's theorem to interchange the integral and the expected value.
This provides the renewal equation for the average infection measure $i(t)$ -- see Eq.~\eqref{eq:i-renewal}.
In the presence of conditioning in Eq.~\eqref{eq:ii-cond}, the right hand side is not linear in $d b$ anymore and we cannot derive an autonomous equation 
for the density $i(t)$ as in the unconditioned case. 
Our approximation now consists in ignoring the non-linearities on the right hand side of Eq.~\eqref{eq:ii-cond} and replacing $d b(t)$ by its expected value.
This gives
\begin{equation}\label{eq:non-linear-approx}
\begin{aligned}
 \widetilde{i}(t) 
\ &= \ \left(\tau(t) \ + \  \int_0^t  \tau(t-u) \widetilde{i}(u) du \right) \ \times \ 
\frac{1-p_{\text{ext}}(0)p_{\text{ext}}(t)\exp\bigg(  \int_{0}^t \ln( p_{\text{ext}}(t-u) ) \widetilde{i}(u) du   \bigg) }{ 1- p_{\text{ext}}(t)\exp\bigg( \int_{0}^t \ln( p_{\text{ext}}(t-u) ) \widetilde{i}(u) du \bigg)  }\\
&= \ F(\widetilde{i}(s); s\in [0,t])\, ,
\end{aligned}
\end{equation}
which explains Eq.~(9) in the main text. 
The overall epidemic size of the conditioned process is then given by $\widetilde{I}(t) = 1 +\int_0^t \widetilde{i}(s)ds$.

\clearpage

\clearpage
\section{Cumulative epidemic size over time}
\label{A:epi_size}
\begin{figure}[h]
    \centering
     \begin{subfigure}{.5\textwidth}
  		\centering
  		\includegraphics[width=\linewidth]{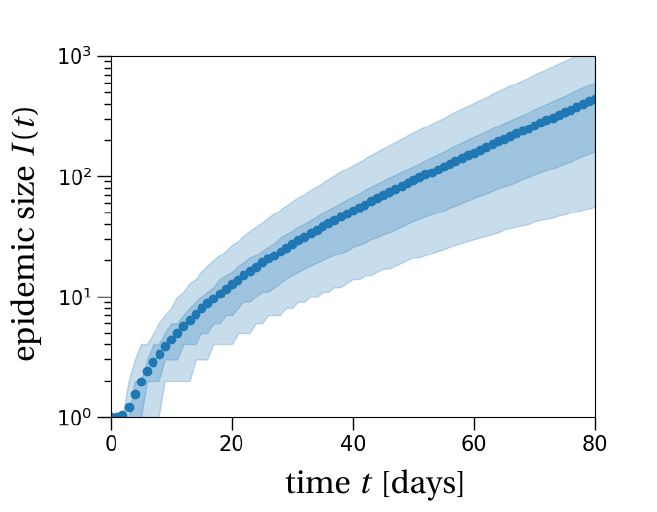}
  		\caption{Poisson}
  		\label{A:fig:Poisson}
	\end{subfigure}%
	\begin{subfigure}{.5\textwidth}
 		 \centering
 		 \includegraphics[width=\linewidth]{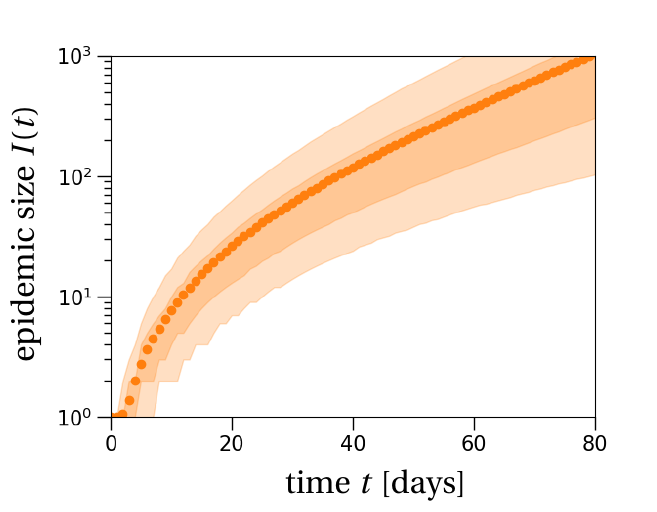}
  		 \caption{negative binomial}
  		 \label{A:fig:negbin}
	\end{subfigure}
	\begin{subfigure}{.5\textwidth}
 		 \centering
 		 \includegraphics[width=\linewidth]{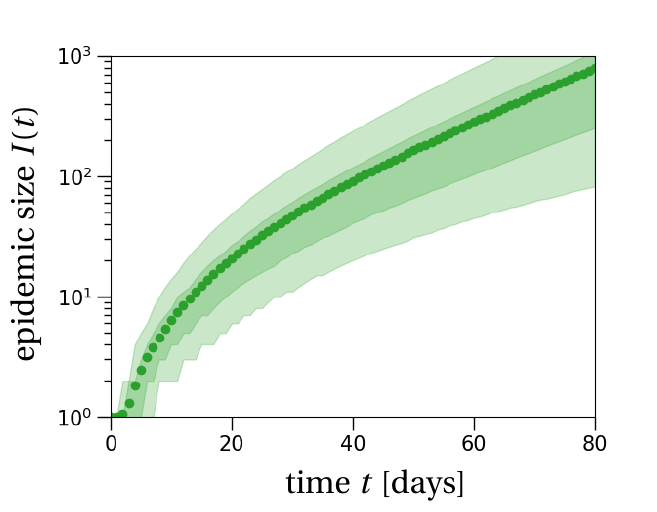}
  		 \caption{geometric}
  		 \label{A:fig:geom}
	\end{subfigure}
  	\caption{\textbf{The cumulative number of infected individuals over time.} The light and dark shaded regions show the 90\% and 50\% inter-quantile ranges obtained from 10,000 stochastic simulations that resulted in establishment of an epidemic cluster. Dots show the average of these simulations over time. With a negative binomial offspring distribution (b), a lot of infected individuals do not transmit the disease, which results in lower epidemic sizes than compared to a Poisson offspring distribution (a). In contrast, few infected individuals with a lot of secondary infections early in the epidemic can generate much larger epidemic sizes when compared to the Poisson distribution. These two effects together explain the much larger variance in epidemic sizes for the negative binomial distribution compared to the Poisson distribution. For a geometric offspring distribution (c), the mean epidemic size and amount of variation are between the numbers found by the negative binomial and Poisson cases. This is explained by the dispersal parameter of the geometric distribution being $\kappa=1$, which is between the respective values for our choice of the negative binomial offspring distribution ($\kappa=0.57$) and the limit $\kappa\to\infty$ corresponding to the Poisson case. The effective reproduction number is set to $R=1.3$ and the transmission density $\mu(t)$ is a Gamma distribution with the parameters as stated in Table~1 in the main text.}
    \label{A:fig:epidemic_size}
\end{figure}

In the main text, we have seen that the cumulative epidemic size over time varies a lot between different stochastic simulations (Fig.~1 in the main text). The number of secondary infections was Poisson-distributed in the main text. The negative binomial and geometric distribution exhibit more variance than a Poisson distribution. Therefore, the variance of the cumulative epidemic size obtained from 10,000 stochastic simulations is even larger in these cases (compare the subfigures in Fig.~\ref{A:fig:epidemic_size}). This can be explained as follows: First, super-spreading events are more likely and as a result, epidemic sizes can be much larger. Secondly, and conversely, a larger number of infected individuals do not transmit the infection at all; as a result, the total epidemic size can remain smaller.

\clearpage
\section{Detection rate of a single mass testing effort}\label{app:mass_test}
\begin{figure}[h]
  	\centering
  	\includegraphics[width=.7\textwidth]{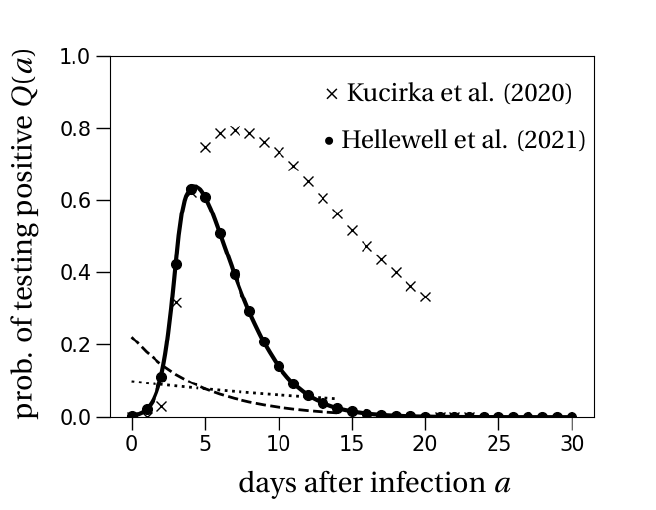}
  	\caption{\textbf{Probability of testing positive.} The probability for an infected individual to test positive by a rapid test (bullet points) and a RT-PCR test (crosses) is close to zero during the first two days after infection. It then increases up to approximately 0.6 on day 4 after infection for rapid tests and up to 0.8 on day 8 for RT-PCR tests. The probability of testing positive then decreases. For rapid tests, the probability of testing positive approaches zero by day 21 after infection. For the RT-PCR test, the probability of testing positive remains relatively high at least until day 20, which corresponds to the last estimated time point in~Fig.~2 (upper panel) from \citet{kucirka2020}. The probabilities for rapid tests are based on data from Fig.~4B in \citet{Hellewell2021}. The dashed and dotted lines show the re-normalized stationary time-since-infection distributions for $R=2.9$ and $R=1.3$ (Eq.~\eqref{eq:q_def}).}
  	\label{fig:detection_proba}
\end{figure}

A good measure to evaluate the prevalence of the disease in a population is the number of detected cases, which depends on the testing effort. In the context of COVID-19, the average detection rate on the 8$^{\text{th}}$ of May 2020 across a large number of countries, mainly in Europe and North America, was inferred to be around 30\% at best \citep{belloir2021,russell2020}. However, test capacity is not the only limiting factor to detect infected individuals. 
The probability for an infected individual to test positive by a reverse polymerase chain reaction or a lateral flow test (LFT), which we will more loosely refer to as rapid tests, varies over the course of the individual's infection \citep{kucirka2020,borremans2020,Hellewell2021}. Here, we will use the data obtained for rapid tests in \citet[Fig. 4B]{Hellewell2021} and for reverse transcriptase polymerase chain reaction (RT-PCR) tests in \citet[Fig. 2]{kucirka2020}, as plotted in Fig.~\ref{fig:detection_proba}. During the first three days of an infection, the viral load within the infected individual is too low to reliably detect the virus with both testing procedures. The probability of detection then reaches a maximum around day four post-infection and then gradually declines. The decline is much faster for rapid tests than for RT-PCR tests. 
In the following, we denote the probability to test positive on day $a$ after infection by $Q(a)$.

Using the probabilities for testing positive by a rapid test and an RT-PCR test (Fig.~\ref{fig:detection_proba}), we compute an upper bound for the fraction of detectable infectious individuals within the epidemic outbreak. Note that this result does not rely on our stochastic correction of the deterministic dynamics, nor does it depend on the epidemic size over time. Instead, the result only depends on the stationary time-since-infection distribution (also called infection age distribution), which is determined by the Malthusian growth rate $r$. Therefore, this result is true for any exponentially growing epidemic. 

Again using general branching process theory \citep{jagers_69,haccou_book}, we find that the number of detectable individuals in the population at time $t$ is given as a fraction $q$ of the overall epidemic size as given in Eq.~(8) in the main text. 
The fraction $q$ is defined as
\begin{equation}\label{eq:q_def}
    q = \frac{1}{C}\  \int_0^{T} Q(a) r e^{-r a} da\, , 
\end{equation}
where $C=\int_0^{T} r e^{-r a}da=1-e^{-T r}$ is the proportion of individuals that have been infected less than $T$ days ago. 
The integral in Eq.~\eqref{eq:q_def} computes the fraction of individuals that are detectable (term $Q(a)$) integrated over the stationary time-since-infection distribution (term $r e^{-r a}$) of the population, restricted to the time since infection being less than $T$ days (constant $C$). 
For the numerical example, we assume that after $T=14$ days infected individuals are not infectious anymore. The resulting truncated distributions for $R=2.9$ and $R=1.3$ are plotted in Fig.~\ref{fig:detection_proba} (dashed and dotted line).

With a high reproduction number $R=2.9$, we find $q\approx 0.255$ (evaluating a discretized version of the integral) when testing is done by rapid tests, so that we would expect that only about $25.5$\% of the infected individuals would test positive with a rapid test at each point in time. Using the RT-PCR data instead, we find $q=0.29$. These estimates strongly depend on the combination of the probability to test positive $Q(a)$ and the stationary time-since-infection distribution. The stationary time-since-infection distribution itself depends on the Malthusian growth rate $r$ (Eq.~(4) in the main text). If we reduce the reproduction number to $R=1.3$, the detection rates become $q=0.265$ and $q=0.48$ for rapid tests and RT-PCR tests, respectively. Interestingly, the fraction of detectable cases does not change in the rapid test scenario substantially, which is a robust result for all possible reproduction numbers between 1.1 and 3. This seems to be due to our choice to only consider individuals with a time since infection less than 14 days. For other choices of this infectiousness threshold, the difference between the detection rates for the two reproduction numbers is more notable. In case of the RT-PCR test, the detection rate increased a lot with a lower reproduction number, from 29\% for $R=2.9$ to 48\% for $R=1.3$. The explanation for this increase in detection rate is that in slowly growing epidemics, more individuals are infected a long time ago and remain detectable with RT-PCR, whereas in faster growing epidemics many individuals are not yet detectable because they were infected within the last two days.

\clearpage
\section{Testing frequency and cluster size at detection}
\begin{figure}[h]
    \centering
    \includegraphics[width=.7\textwidth]{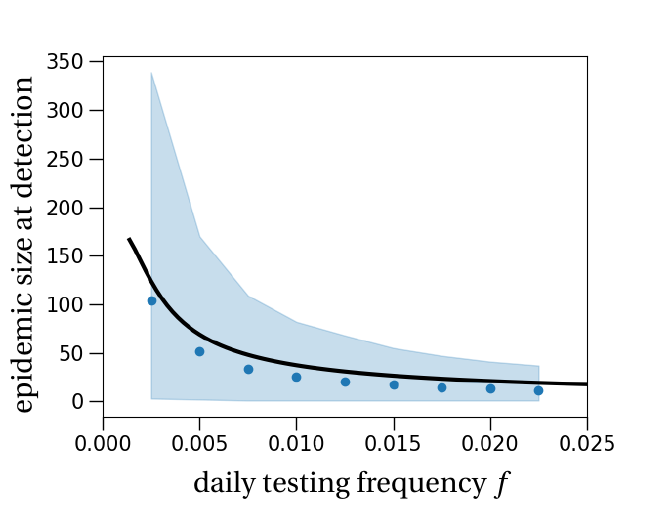}
  	\caption{\textbf{Average epidemic cluster size at detection depending on the daily testing frequency.} The shaded region shows the 90\% confidence interval of the cluster size obtained from 10,000 stochastic simulations that resulted in cluster establishment. Dots represent the average cluster sizes of these simulations at the first detection time. The theoretical prediction (black solid line) is obtained from Eq.~(11) in the main text, adapted to the setting of testing. In particular, the distribution until detection is motivated by the probability to test positive in Fig.~S3 in SI, Section~S7, and translated to a detection probability by Eq.~(12) in the main text. The effective reproduction number is set to $R=1.1$ and the number of secondary infections is Poisson-distributed. The transmission density $\mu(t)$ is as stated in Table~1 in the main text.}
    \label{fig:testing}
\end{figure}

We study the average epidemic cluster size at detection when varying the daily testing frequency $f$ in a population. Fig.~\ref{fig:testing} shows that our analytical prediction based on Eqs. (11) and (12) from the main text slightly overestimates the averages as obtained from stochastic simulations. Importantly, the epidemic size at detection reflects the exponential growth of the epidemic and declines exponentially with an increasing daily testing frequency. This indicates that there might be a clear optimal testing frequency that ensures clusters are relatively small when detected while not overwhelming test systems.  

\end{appendix}

\end{document}